\documentclass[sigconf]{acmart}

\usepackage{booktabs} 

\usepackage{pifont}
\usepackage[all]{hypcap}
\usepackage{amssymb}
\usepackage{amsmath}
\usepackage{hyperref}

\usepackage{color, colortbl}
\definecolor{Gray}{gray}{0.9}

\usepackage{acronym}
\newacro{iot}[IoT]{Internet of Things}
\newacro{iiot}[IIoT]{Industrial Internet of Things}
\newacro{manet}[MANET]{Mobile Ad Hoc Network}
\newacro{pin}[PIN]{Personal Identification Number}
\newacro{otp}[OTP]{One Time Password}
\newacro{ip}[IP]{Internet Protocol}
\newacro{cots}[COTS]{Commercial Off The Shelve}
\newacro{mac}[MAC]{Mandatory Access Control}
\newacro{dac}[DAC]{Discretionary Access Control}
\newacro{rbac}[RBAC]{Role-Based Access Control}	
\newacro{radius}[RADIUS]{Remote Authentication Dial-In User Service}
\newacro{rfid}[RFID]{Radio Frequency IDentification}
\newacro{sdn}[SDN]{Software-Defined Networking}
\newacro{tsn}[TSN]{Time-Sensitive Networking}
\newacro{nfv}[NFV]{Network Function Virtualisation}
\newacro{hp}[HP]{Honeypot}
\newacro{hn}[HN]{Honeynet}
\newacro{scada}[SCADA]{Supervisory Control and Data Acquisition}
\newacro{opcua}[OPC-UA]{OPC-Unified Architecture}
\newacro{mes}[MES]{Manufacturing Execution System}
\newacro{erp}[ERP]{Enterprise Resource Planning}
\newacro{hp}[HP]{Honeypot}
\newacro{plc}[PLC]{Programmable Logic Controller}
\newacro{siem}[SIEM]{Security Information and Event Management}
\newacro{ids}[IDS]{Intrusion Detection System}
\newacro{nic}[NIC]{Network Interface Card}
\newacro{wsn}[WSN]{Wireless Sensor Network}

\begin{document}
\title{The Dos and Don'ts of Industrial Network Simulation: A Field Report}

\author{Simon Duque Anton}
\affiliation{%
  \institution{Intelligent Networks Research Group\\German Research Center for Artificial Intelligence}
  \city{Kaiserslautern}
  \country{Germany}
  \postcode{67663}
}
\email{simon.duque_anton@dfki.de}

\author{Daniel Fraunholz}
\affiliation{%
  \institution{Intelligent Networks Research Group\\German Research Center for Artificial Intelligence}
  \city{Kaiserslautern}
  \country{Germany}
  \postcode{67663}
}
\email{daniel.fraunholz@dfki.de}

\author{Dennis Krummacker}
\affiliation{%
  \institution{Intelligent Networks Research Group\\German Research Center for Artificial Intelligence}
  \city{Kaiserslautern}
  \country{Germany}
  \postcode{67663}
}
\email{dennis.krummacker@dfki.de}

\author{Christoph Fischer}
\affiliation{%
  \institution{Intelligent Networks Research Group\\German Research Center for Artificial Intelligence}
  \city{Kaiserslautern}
  \country{Germany}
  \postcode{67663}
}
\email{christoph.fischer@dfki.de}

\author{Michael Karrenbauer}
\affiliation{%
  \institution{Institute for Wireless Communication and Navigation\\Technical University of Kaiserslautern}
  \city{Kaiserslautern}
  \country{Germany}
  \postcode{67663}
}
\email{karrenbauer@eit.uni-kl.de}

\author{Hans D. Schotten}
\affiliation{%
  \institution{Intelligent Networks Research Group\\German Research Center for Artificial Intelligence}
  \city{Kaiserslautern}
  \country{Germany}
  \postcode{67663}
}
\email{schotten@dfki.de}

\renewcommand{\shortauthors}{S. Duque Anton et al.}

\settopmatter{printacmref=false}

\copyrightyear{2018}
\acmYear{2018}
\setcopyright{acmlicensed}
\acmConference[ISCSIC '18]{The 2nd International Symposium on Computer Science 
and Intelligent Control}{September 21--23, 2018}{Stockholm, Sweden}
\acmBooktitle{The 2nd International Symposium on Computer Science and 
Intelligent Control (ISCSIC '18), September 21--23, 2018, Stockholm, Sweden}
\acmPrice{15.00}
\acmDOI{10.1145/3284557.3284716}
\acmISBN{978-1-4503-6628-1/18/09}

\begin{abstract}
Advances in industrial control lead to increasing incorporation of intercommunication technologies and embedded devices into the production environment.
In addition to that,
the rising complexity of automation tasks creates demand for extensive solutions.
Standardised protocols and commercial off the shelf devices aid in providing these solutions.
Still,
setting up industrial communication networks is a tedious and high effort task.
This justifies the need for simulation environments in the industrial context,
as they provide cost-,
resource- and time-efficient evaluation of solution approaches.
In this work,
industrial use cases are identified and the according requirements are derived.
Furthermore,
available simulation and emulation tools are analysed.
They are mapped onto the requirements of industrial applications,
so that an expressive assignment of solutions to application domains is given.\\ \par
This is a preprint of a work published in the Proceedings of the 2nd International Symposium on Computer Science and Intelligent Control (ISCSIC 2018).
Please cite as follows: \par
S. D. Duque Anton, D. Fraunholz, D. Krummacker, C. Fischer, M. Karrenbauer, and H. D. Schotten, : ``The Dos and Don'ts of Industrial Network Simulation: A Field Report.''
In: Proceedings of the 2nd International Symposium on Computer Science and Intelligent Control (ISCSIC 2018), 2018, pp. 6-13.
\end{abstract}

%
%



\maketitle

\section{Introduction}\label{sec:intro}
Industrial networks have been proprietary systems,
set up for each specific application scenario.
With the rising importance of the \ac{iiot},
standardised and extensive interconnectivity gains importance.
The influence of networking in production is increasing,
control and maintenance tasks can be performed remotely \cite{karrenbauer2018}.
Well-established communication protocols,
such as \textit{EtherCAT},
\textit{Modbus},
\textit{Modbus/TCP},
\textit{PROFINET}
and \ac{opcua},
as well as \ac{cots} hardware reduce the configuration and customisation effort in setting up networks.
On the other hand,
the increasing integration of communication into automation creates a deeper penetration of industrial applications by communication and computational devices.
Recently,
wireless protocols,
such as \textit{WirelessHART},
\textit{ZigBee} and \textit{LoRa} add to the variability of solutions.
The increasing importance of networking for industrial applications justifies a rising need for network simulation tools of industrial environment.
It is necessary for planning of network structures,
for the evaluation of timing criteria,
for adaption of control algorithms in \ac{sdn},
and also for security applications.
This work aims at providing an overview of existing methods for simulation and emulation,
enabling users to chose the best suited method and to create the best fitted environment.
\\
The remainder of this work is structured as follows.
In section~\ref{sec:sota},
an overview of works analysing and categorising simulation environments is provided.
After that,
relevant uses cases for simulation of industrial networks are introduced in section~\ref{sec:uc}.
Requirements derived from these use cases are explained in section~\ref{sec:req},
solution approaches are presented in section~\ref{sec:solution} and a mapping of these solutions to the requirements is provided in section~\ref{sec:sol_map}.
The work is concluded in section~\ref{sec:conclusion}.

\section{Related Work}
\label{sec:sota}
Due to the novelty of this topic,
not much work has been done to evaluate simulation tools for industrial applications.
\textit{Terzi and Cavalieri} considered simulation as a part of supply chain management and a tool for change management~\cite{Terzi.2004}.
Apart from that,
work has been done in evaluating well-established simulation tools for their performance.
\textit{Weingartner et al.} compare the performance of five network simulators~\cite{Weingartner.2009},
while \textit{Flores Lucio et al.} analyse accuracy of \textit{ns-2} and \textit{OPNET}~\cite{Flores_lucio.2003}.
\textit{Pan}~\cite{Pan.2008},
as well as \textit{Siraj et al.}~\cite{Siraj.2012} consider simulator environments and testbeds as means to save money, time and effort in network planning.
They present several tools and compare advantages and disadvantages.
The work of \textit{G\"{o}kt\"{u}rk} addresses the same topics,
after systematically distinguishing between testbeds and emulation~\cite{Gokturk.2007}.
Furthermore,
there are surveys addressing specialised fields of network simulation.
\textit{K\"{o}sal} analyses four well-known simulators that support wireless networks~\cite{Koksal.2008},
\textit{Naicken et al.} consider peer-to-peer network simulators~\cite{Naicken.2006},
\textit{Martinez et al.} address vehicular ad hoc networks~\cite{Martinez.2011},
and the work of \textit{Sarkar et al.} is about simulation of telecommunication networks~\cite{Sarkar.2011}.
In addition to these use cases,
several works address simulation tools for \acp{wsn}~\cite{Sundani.2011, Imran.2010, Korkalainen.2009, Christin.2010, Yu.2011}.
\acp{wsn} are becoming more relevant with the increase in embedded computing power.
Distributing \acp{wsn} contains a relatively large organisational overhead,
justifying the use of sound simulation environments.

\section{Use Cases for Industrial Network Simulation}
\label{sec:uc}

In this section,
the different use case scenarios motivating the appliaction of simulation environments are described in subsection~\ref{ssec:ucs}.
Developments in technologies of interconnectivity,
as well as the integration of \ac{iot} devices into the industrial environment create novel types of industrial networks,
e.g. the \ac{iiot}.
After that,
the simulation scopes are described in subsection~\ref{ssec:scope}.
Some use cases require a high-level view of a network without the need for simulating a detailed behaviour,
while others need well-defined traffic packets.
Finally,
the difference between simulation and emulation is explained in subsection~\ref{ssec:simvsem}.

\subsection{Use Case Scenarios}
\label{ssec:ucs}



In this subsection, 
the prototypical scenarios of industrial use cases that could benefit from simulation are described.

\subsubsection{Network Management Algorithms}
A new set of use cases is introduced due to Industry 4.0~\cite{3gpp2017},
entailing new challenges to industrial communication networks.
The derivation of the
challenges are introduced in \cite{DenKrChFiAR2018}.
These challenges create demand for a sophisticated network management that,
on one hand,
is capable of handling a heterogenous landscape of industrial communication protocols.
On the other hand,
it needs to satisfy requirements like ultra low latency,
determinism,
cross domain communication,
dependability,
flexibility and security.
There are innovative technologies at hand that meet these requirements,
sometimes in conjunction with one another.
Examples for these technologies are \ac{sdn},
which provides flexibility and scalability to networks,
and \ac{tsn},
which allows for determinism of the network,
as well as \ac{nfv},
which can control dependability relevant parameters in a centralized management system.
The modular combination
of technologies can cause interdependency,
inferring the danger of a live-lock behaviour
in a control or management system.
Due to this issue,
sophisticated algorithms capable of combine technologies and resolve interdependency are required.
Network management is needed in several points of the work flow.
First of all, integration of \ac{tsn} and \ac{sdn} into a network can not be performed on a single switch.
Complex network scenarios have to be set up
to be developed and evaluated iteratively.
Algorithms for network management require these underlying technologies in order to be developed for a network.
The above mentioned circumstances yield to sophisticated requirements on the test environment in terms of flexibility, openess of interfaces and scalability. 
Due to this a simulation is a preferable way to deploy the test environment.

\subsubsection{Digital Factory}
The \textit{Digital Factory} is a use case described by \textit{Stef at al.}~\cite{STEF2013451}.
It focuses on the simulation of a production unit in an industrial environment.
In doing so,
the debugging can be executed in advance of the setup and the
deployment of the production unit can be accelerated. Besides the simulation of the production unit, a simulation of the network is crucial to provide an appropriate dimensioned network capable of meeting all technological
demands. Furthermore, it is possible to deploy major changes in network settings in a simulated environment and run through the debugging process, before implementing it on the running hardware.
In this way, possible network crashes can be avoided. In this case the advantages of a simulation concerning deployment speed, cost reduction are the main arguments for their usage.
One characteristic of the digital factory is a flattened automation pyramid. Control is not necessarily introduced from a higher layer anymore.
Instead, entities on the same layer may communicate through network boundaries. Many of such use cases include the necessity for real-time communication, that consequently has to be addressed by the simulation tool chosen for the scenario.

\subsubsection{Honeypots}
Deception is a well established method in many domains.
In IT security it is known as \textit{Deception Technology}.
The most noteable implementation are \acp{hp}.
\acp{hp},
in their very essence,
mimic resources such as \ac{plc}s or \ac{scada} systems.
The deceptive capabilites rely on the quality of simulation,
more precise and comprehensive imitations allow for a deeper insight into the interaction and therefore the motivation,
expertise and arsenal of hostile entities.
As these insights are valueable for prevention \cite{Fraunholz.2018b},
detection \cite{Fraunholz.2018} and attribution \cite{Fraunholz.2017e,Fraunholz.2017d,Fraunholz.2017f},
the use of \ac{hp}s is promising for industrial security.
They are also able to cope with the most pressing challenges,
such as durability,
latency,
heterogeneity and encryption.
The simplest industrial \ac{hp}s support common industrial protocols,
such as \textit{Bacnet},
\ac{opcua} or \textit{modbus} \cite{Rist.2015}.
For those protocols,
that are frequently employed in enterprise and industrial environments,
a large selection of \ac{hp}s is available.
Examples are HTTP/HTTPS~\cite{Provos.2004},
FTP \cite{CarnivoreProject.2009} and Telnet/SSH~\cite{Oosterhof.2014}.
Since productive systems are usually interconnected,
so are \acp{hp}.
In this case,
they are referred to as \acp{hn}.
The infamous \textit{honeyd} \cite{Provos.2004} is an example for a \ac{hn} implementation.
Besides communication capabilites,
system simulation is also an integral part of deception.
Recent implementations use \textit{IMUNES}/\textit{Conpot} \cite{Kuman.2017},
\textit{Matlab}/\textit{Simulink} \cite{Litchfield.2017} or \textit{gridlabd} \cite{GridPot:SymbolicCyberPhysicalHoneynetFramework.2015, Redwood.2015} for system simulation.

\subsubsection{Industrial IT Security}
The increasing interconnectivity and rising connection of industrial to public networks leads to an enlarged attack surface of these networks~\cite{Duque_Anton.2017a}.
In contrast to home and office IT, 
industrial networks have been set up with little thought of security.
Since this resulted in,
sometimes drastic,
attacks in recent years,
this paradigm is changing.
Security solutions,
such as industrial firewalls and \ac{siem} systems are being developed.
Furthermore,
there is a large research interest in detecting anomalies and intrusions in industrial network traffic.
Context information has proven to be a valuable indicator~\cite{Duque_Anton.2017c},
but is not trivial to obtain~\cite{Duque_Anton.2017b}
Apart from IT security reasons,
anomalies can be an indicator of misconfiguration or malfunction,
motivating the need to detect them in a timely manner.
There are some works regarding anomaly and intrusion detection in industrial network data~\cite{Duque_Anton.2018}.
Several surveys cover this topic,
such as the work of \textit{Garcia-Teodoro et al.}~\cite{Garcia-Teodoro.2009} and \textit{Bhuyan et al.}~\cite{Bhuyan.2014}.
\textit{Yang et al.} give a brief introduction of these techniques for the domain of \ac{scada} systems~\cite{Yang.2006}.
\textit{Meshram and Haas} published a roadmap of machine learning based anomaly detection in industrial networks,
containing a simulation environment,
as well as a semantic description of content~\cite{Meshram.2016}.
\textit{Kleinman and Wool} present a model of the \textit{Siemens S7} protocol for intrusion detection and forensics~\cite{Kleinmann.2014}.
Critical infrastructures and industrial environments are considered in the work of \textit{Hadziosmanovic et al.}~\cite{Hadziosmanovic.2011}.
A framework that detects malicious and undesired actions is presented.
Deriving features that can be used to distinguish valid from malicious traffic is the first step in applying an intrusion detection algorithm.
\textit{Mantere et al.} look into the derivation of features from \ac{ip} traffic in an industrial environment~\cite{Mantere.2013}.
Deterministic properties of industrial control systems,
as well as the usability of this feature for anomaly detection in an industrial environment,
is researched by \textit{Hadeli et al.}~\cite{Hadeli.2009}.
Most of these algorithms need a large amount of data in order to be properly trained.
In contrast to home and office network traffic,
industrial network traffic is still rare.
This is an issue \textit{Morris and Gao} describe in their work~\cite{Morris.2014}.
Therefore,
they introduce an intrusion detection dataset that,
however,
consists of aggregated data instead of network packets.
This renders is unfit for anomaly detection algorithms that are packet- or timing-based.\\
In general,
due to concerns regarding privacy,
confidentiality and effort,
simulation will play a crucial role in intrusion detection for industrial systems.
Real world data sets are difficult to obtain,
mostly due to organisational issues.
This means that research has to use synthetic data or data that has been created in research environment.
The realism and consistency of this data is crucial,
as the results of an industrial \ac{ids} depend on them.
Flawed data could lead to good results on the test data,
but weak performance in a productive environment.
As the demand and necessity of industrial intrusion detection will increase over the next years,
the need for data will rise.

\subsection{Scope of Simulation}
\label{ssec:scope}
Apart from the use case scenarios as described above,
the scope of the simulation is of interest.
First, 
the scope in terms of automation level is discussed.
After that,
the scope in terms of geographical spreading is examined.

\subsubsection{Level of Automation}
\begin{figure}[!ht]
\centering
\includegraphics[width=0.30\textwidth]{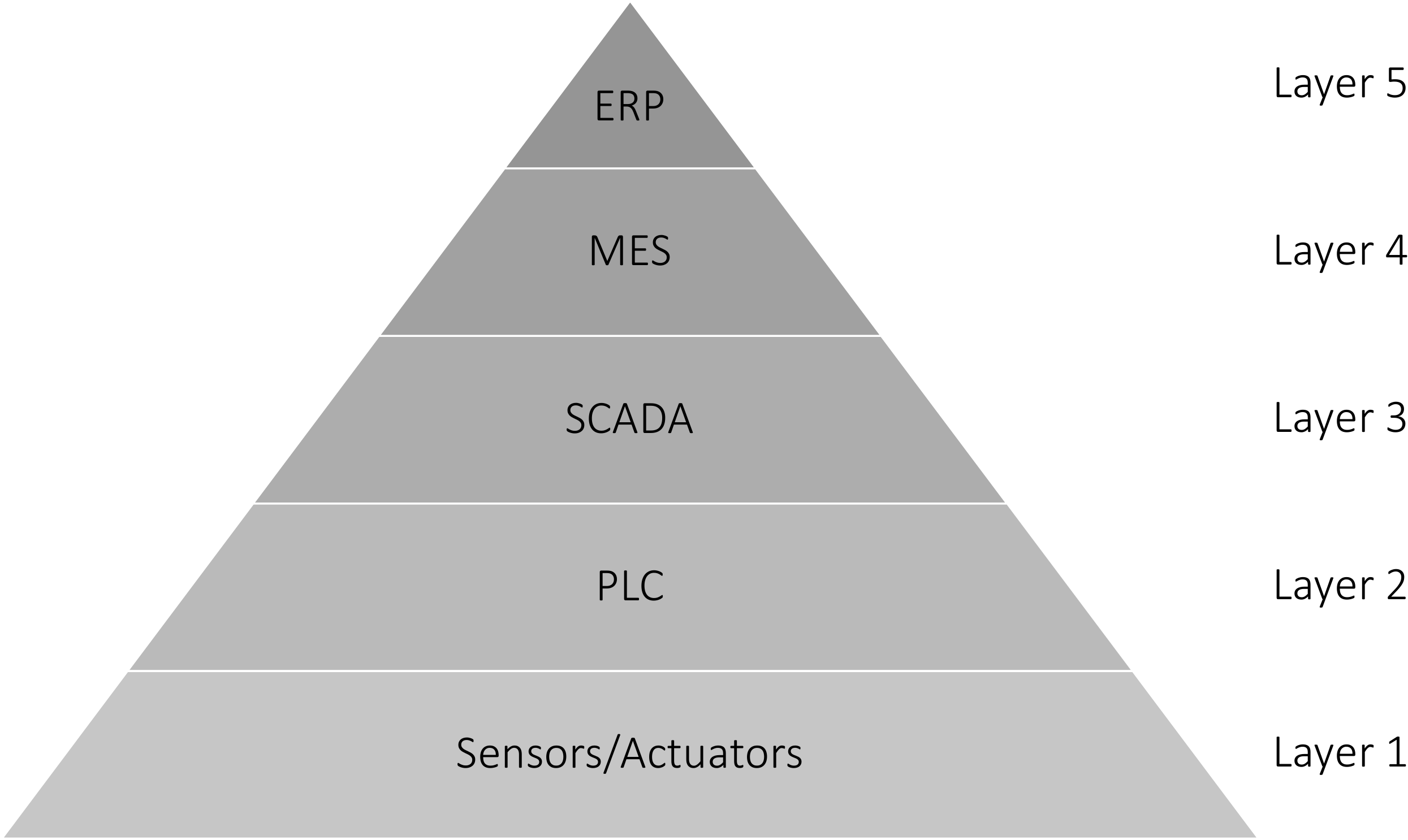}
\caption{Automation Pyramid}
\label{fig:automation_pyramid}
\end{figure}

In the 1990's,
so-called automation pyramid has been introduced.
It is depicted schematically in figure~\ref{fig:automation_pyramid}.
The different levels of a hierarchical industrial network are explained with this automation pyramid,
as there are different requirements for different levels.
On layers 1 and 2,
for example,
real time communication is often required.
In addition to that,
a physical model of the production system is required for layer 1,
which is usually difficult to implement.
Layers 4 and 5,
\ac{erp} and \ac{mes};
systems,
that control the setup and production processes of automation devices,
on the other hand,
are mostly implemented using standard TCP/IP networks.
The \ac{scada} system on layer 3 commonly employs non-real time \textit{Ethernet}-based protocols.

\subsubsection{Geographical Scope}
\begin{figure}[!ht]
\centering
\includegraphics[width=0.48\textwidth]{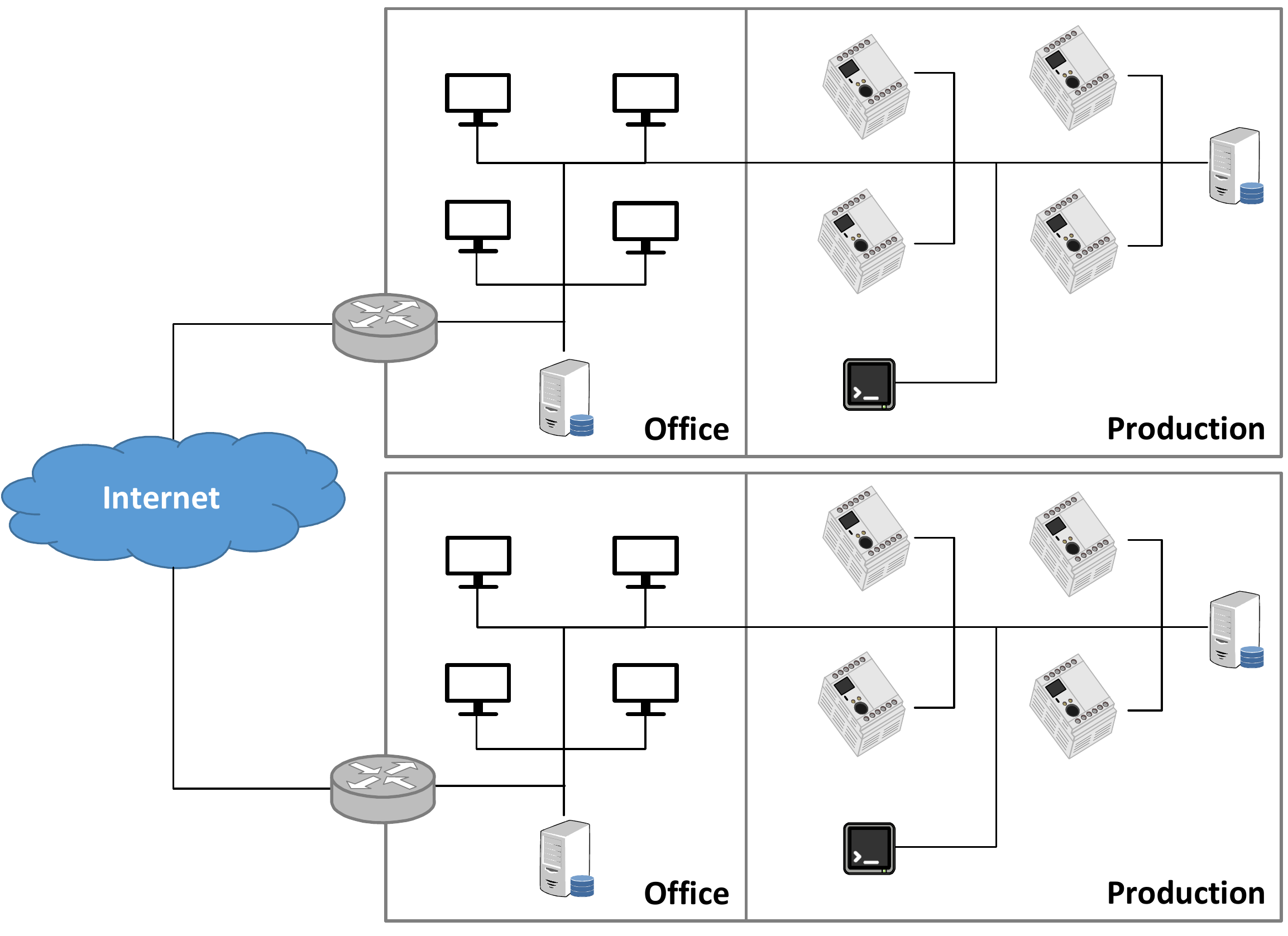} 
\caption{Geographical Scope}
\label{fig:phys_scope}
\end{figure}

Per Se,
networks are employed to connect different systems and enable communication.
There are several characteristics of networks.
A typical set of scenarios is depicted in figure~\ref{fig:phys_scope}.
Depending on the scenario,
simulation of traffic in different parts of the shown networks can be relevant.
A single connection between the \acp{plc} in the \textit{production} part of the figure can be of interest,
e.g. when the characteristic of a certain antenna is to be determined.
For network management,
a whole site, 
including connection between \textit{office} and \textit{production} area can be under investigation.
Furthermore,
the connectivity possibilities over public networks can be analysed,
e.g. for security analysis.

\subsection{Simulation vs. Emulation}
\label{ssec:simvsem}
In order to categorise solutions,
a terminology has to be introduced first.
Generally,
there is a semantic difference between simulation and emulation,
even though these terms are often used interchangeably.
In this work,
we will generally use the term ``simulation'' for the sake of readability,
since many tools can be used for both.
\\ \par
Generally,
simulation aims at mimicking the states of a system~\cite{Cellier.1991},
while emulation aims at reproducing the behaviour~\cite{KB.}.
This difference often becomes relevant in the type of experimental setup:
As discussed in the following subsection,
simulation is usually done on one computer.
That means that any physical characteristic has to be modeled.
The goal is to create a model of the components under investigation and imitate the behaviour,
as well as internal states as accurately as possible.
Sometimes,
this leads to a high effort in creating the simulation model.
Emulation,
on the other hand,
can be done on real network infrastructure with devices imitating other devices.
If only the transmission behaviour is of interest,
then the nature of devices creating the traffic is irrelevant. 
Emulation therefore has the advantage that its output is,
by design,
as accurate as the productive system,
since the physical environment is used.
On the downside,
it usually has a higher requirement in devices and physical space.

\section{Requirements for Industrial Network Simulation}
\label{sec:req}
In this section, each of the use case scenarios, as described in subsection~\ref{ssec:ucs}, is analysed. The requirements on a simulation or emulation environment are derived from the characteristics and explained. Furthermore, the scope in terms of abstractivity and geographical extension, as described in subsection~\ref{ssec:scope}, is considered.

\subsection{Network Management Algorithms \& Digital Factory}

Industrial production is gaining flexibility. Future industrial production systems are getting an increasingly dense interconnectivity that has to serve a rising amount of network traffic,
inflicting stringent requirements on the network.
This raises the necessity of an autonomous operating and automatic self-organising-network to be able to master the complexity. A possible solution for a Network Management Concept is presented in \cite{DenKrChFiAR2018}.
A Simulation has to mimic the real-world behaviour as accurately as possible, when it comes to metering especially performance dependent aspects of such systems. Even more important is an accurate replica of industrial network systems for development of new solutions.
In case of analysing the fundamental behaviour of transmissions on a network instead of functional high-level solutions,
accuracy of the simulation is crucial.
The scope of this simulation sometimes extends to the characteristics of singular packets on the physical transmission medium,
e.g. in safety packets that inflict hard real-time requirements.

\subsubsection{Timing Precision}
A key demand for digital factories is the ability to grant certain transmission guarantees,
e.g. maximum transmission time or reliability.
While developing new transmission mechanisms,
protocols or overlaying methods influencing transmission procedures,
it is important to have the ability to measure the timing behaviour of transmissions under the given circumstances.
The essential point is that not only the functional aspect of macroscopic or cumulated operations has to be evaluated,
but rather the fine-grained runtime performance of a set of operations.
This requires the simulation of a highly precise and representative time base on which a realistc imitation of the transmission time behaviour can be achieved.
For protocols and low level methods,
it has to be evaluated if they grant the ability to meet requirements,
for instance transmission deadlines in general.
Higher level methods need to prove that they properly interact with underlying or cooperating procedures without disrupting.

\subsubsection{Flexibility / Extensibility}
A huge variety of differing communication technologies, protocols and vendor-specific proprietary solutions is assembled in industrial networks.
In order to address this characteristic by a simulation tool,
it must provide the possibility to integrate novel modules above the physical level,
such as protocols or medium access methods.
Additionally,
when working on new communication technologies or fundamentals of them,
e.g. new waveforms of wireless transmissions,
the physical and transmission models of the simulation have to be adapted.
For example,
the introduction of differing channel models to adjust the simulation to different environments can be of value.

\subsubsection{Real-World Interaction}
Often,
existing parts of a network infrastructure should be integrated into a simulation.
This can be due to the extension of a network,
or the effort it would take to create a model of devices or infrastructure already existing.
Basically,
network simulation is predestined to interact with external,
real-world domains.
In order to efficiently do so,
simulation environments need to provide interfaces to communicate with protocols and devices they are modelling.


\subsubsection{Trusty Physical Imitation}
The ability to accurately measure physical influences on networks,
to enable a realistic comparison of different communication technologies,
is of high value.
These physical influences need to be modelled with sufficient accuracy for the simulation to provide results that will not deviate from any real-world application.
Usually,
not all parameters are of equal importance.
An application example is the comparison of fading effects in narrow-band and ultra-wide-band wireless communication.

\subsection{Honeypots}
In contrast to other use cases,
\ac{hp}s only rely on the perception of the entity that it intends to deceive \cite{Fraunholz.2017g}.
It is therefore important to define the attack vector and simulate the corresponding observations a hostile entity may encounter.
For example, if the attack scenario is most likely network sniffing, the network packets should be generated in accordance to the communication stack of the system that is simulated. From a technical perspective, there is a trade-off between quality of simulation and handling of the systems. Systems with a more realistic imitation tend to be more complex to monitor and secure as there is a wider variety of possible actions.
There is an abundance of examples, where systems that provide full-fledged operating systems are suffering from easy detection, relay attacks and monitoring circumvention \cite{Fu.2006,Chen.2008,Mukkamala.2007,Dornseif.2004,Holz.2005}.
To mitigate detection attempts, it is of significant importance to enable a dynamic configuration, deployment and management of such systems \cite{Fraunholz.2017b,Fraunholz.2017}.
However, not only technical requirements need to be considered when applying such technologies.
It must be ensured that the application is conform with law of the corresponding jurisdiction \cite{Fraunholz.2017i}.

\subsection{Industrial IT Security}
Industrial IT security is a very broad area with an abundance of requirements.
In this work,
an effort is undertaken to generalise them while being accurate enough to derive possible solutions.

\subsubsection{Injectability of Malware}
The first and foremost requirement of industrial network data generation for IT security is the possibility to inject malicious traffic.
This malicious traffic has to fit the environment and correctly represent the instantiation of a possible attack vector.
In emulation environments with real network infrastructure,
executing malware with common penetration testing tools,
such as \textit{metasploit},
is an option.
This,
however,
is not possible in network simulations running on one machine.
If simulation is chosen,
the malicious behaviour needs to be modeled and injected in a valid and sound fashion.

\subsubsection{Variability}
Due to the inherent unique nature of most industrial networks,
a versatile and adaptive data generation environment is requried.
In simulation,
as well as emulation,
adapting or creating new scenarios requires reconfiguration and possibly creation of new models.
Tools need to be able to adapt to these changes.

\subsubsection{Viability and Soundness}
Especially when employing methods of machine learning,
many algorithms are prone to overfitting.
That means they focus on parameters that have little to nothing to do with the actual traffic,
but are artefacts of data generation.
The DARPA KDD cup '99,
for example,
contains one feature that is specific to all malicious instances,
but solely stems from the simulation model.
If \acp{ids} are tested on such data,
they lack performance once they are integrated into real world processes.
To prevent focussing on invalid features and overfitting,
data generation has to be sufficiently sound and compatible to real world data.
This is commonly considered as the most difficult task in data generation,
especially since a user often cannot determine which feature has which influence on an algorithm.

\subsection{Requirements Summary}
\label{ssec:requirements_summary}
So far in this section,
the requirements on all of presented use cases have been described.
In order to map the possible solutions,
as described in section~\ref{sec:solution},
the requirements are summarised.
In general,
there are five requirements for industrial network simulation that are listed below:
\begin{itemize}
\item Timing precision describes the accuracy and timeliness of traffic,
which needs to be highly precise for \ac{tsn}.
\item Flexibility in setting up and changing simulation characteristics
\item Real world interaction, the capability to be integrated into existing productive environments
\item Trusty (physical) imitation describes the soundness and correctness of simulation
\item Adaptability describes the potential to be adjusted and mapped to novel scenarios
\end{itemize}

\section{Solution approaches}
\label{sec:solution}
In this section,
several solutions are proposed and their advantages and disadvantages discussed.
The solutions presented in this section are:
Application specific solutions,
\textit{ns-3},
\textit{OMNeT++},
and \textit{Mininet}.

\subsection{S1: Application Specific Solutions}
Creating an application specific simulation tool allows for the highest level of flexibility,
on the cost of the most effort needed.
They can be either simulation or emulation environments,
or a mix.
Furthermore,
real components can be used to model the system.
Real network parts can be used with simulated devices,
for example.
For industrial IT security,
\textit{Lemay and Fernandez} provided a data set from an application specific simulation~\cite{Lemay.2016}.
\textit{Wang et al.} provide a simulation environment for \ac{scada} security assessment~\cite{Wang.2010},
as does \textit{Seidl}~\cite{Seidl.2015}.
and \textit{Siaterlis et al.}~\cite{Siaterlis.2013},
as well as \textit{Genge et al.}~\cite{Genge.2012} introduce testbeds for research in industrial IT security.

\subsection{S2: ns-3}
\textit{Ns-3} is the name of an event-based open source network simulator which can be used for researching on IP and non-IP based networks. It aims at providing simulation models that are as realistic as possible and can be used as a network emulator, i.e. interconnected with real world devices. \textit{Ns-3}, being an event-based simulator, already provides a time base and many other modules, including a WiFi PHY and MAC layer implementation, wireless channel models and protocol stacks. Therefore, \textit{ns-3} is a sound choice for simulating the inter-connectivity of different networks, especially but not limited to wireless networks, which makes it appropriate for modeling the use case scenarios ''Network Management Algorithms`` and ''Digital Factory``.
\textit{Ns-3} provides different simulation models.
In a closed simulation, \textit{ns-3} traverses through the simulation whilst it maintains a changing event-list. Throughout runtime,
\textit{ns-3} processes one event after the other until the event-list is empty or a specified point in time for the simulation end is reached. Events are caused explicitly by instructions from the simulation program or can be raised by other events during runtime (possibly infinitely).
Alternatively \textit{ns-3} supports a real-time scheduler that facilitates interaction with real systems.

\textit{Ns-3} itself is written in C++ and for its use provides libraries for the programming language C++. Accordingly, \textit{ns-3} simulations are in fact programs, written in C++, using the \textit{ns-3} libraries. An \textit{ns-3} simulation defines the involved network devices and per device the used network stack and eventually running applications. Between the devices the communication technology is specified with an explicitly chosen channel model. If desired,
one could implement custom channel models, completely new underlying calculation methods or new physical influences and thereby introduce new technologies. Admittedly, doing so would require a high effort.
The out-of-the-box implemented transmission channel replications of \textit{ns-3} move somewhere between realistic accuracy and low computational expense but are adequately representative, like evaluated by \cite{boeingNS3ChanEval}.


Likewise \textit{ns-3} can be extended with new protocols or any other module that incorporates somewhere in the processing of data through a network stack, across a network or if desired completely local on one machine.
Network devices in \textit{ns-3} aren't full-featured devices inclusive operating system (like it would be in the case of mininet).
Rather every cooperating piece has to be specified. As its purpose to be a network simulator, the most common networking features are already available in \textit{ns-3} to be deployed to simulated devices.
It is a substantial property of \textit{ns-3} that it is open source and high-grade modular, which creates a maximum on flexibility with the choice of willingness for effort up to the developer of a simulation. Many features and accuracy are already available in \textit{ns-3}. If one desired is not, it is likely that the implementation of it would be possible with a smooth incorporation to distinct \textit{ns-3} simulation modules.
\textit{Ns-3} indeed creates network packets and passes them like real devices would. As a consequence,
these packets can be treated like real ones, for instance can be sniffed using libpcap, analysed, logged, malformed or modified. \textit{Ns-3} offers the ability to capture and process packets \textit{live}, i.e. during the running simulation or to log them for later use in various formats,
like .pcap or plain ASCII.


\subsection{S3: OMNeT++}
The \textit{\textbf{O}bjective \textbf{M}odular \textbf{Ne}twork \textbf{T}estbed in C\textbf{++} (OMNeT++)}~\cite{OMNeT++.2015} provides a C++-based network simulation platform,
even though ``\textit{OMNeT++} is not a network simulator itself''.
Devices and networks are described within the environment provided.
They are then connected and communicate according to a given pattern on the physical layer,
based on a discrete event simulator.
Integration of higher layers is possible by employing so-called frameworks,
that implement protocols and functionality and can be used by the \textit{OMNeT++} environment.
Due to this extendibility,
any functionality can be integrated into the \textit{OMNeT++} environment.
Wireless,
as well as real time applications can be simulated.
Furthermore,
the use of \textit{OMNeT++} as an emulation testbed over a real network is possible by employing sockets.
In general,
the functionality is similar to \textit{ns-3}.
There are graphical user interfaces available for \textit{OMNeT++},
making introduction easier.
Furthermore,
the licensing model is different:
While there is an open source license for academia,
industrial users have to purchase a proprietary \textit{OMNeT++} distribution.

\subsection{S4: Mininet}

\textit{Mininet} \cite{mininetWeb} is a network emulator available for Linux Operating Systems. It is more high-level and oriented on using a network (e.g. for development or testing of applications or control algorithms) instead of researching on low-level network topics itself (such as channel access schemes or synchronisation protocols for instance).
On boot,
it creates a live interactive emulation environment, using Linux Kernel utilities. \textit{Mininet} creates a network of virtual hosts, switches, controllers and links, within whose the hosts run standard Linux network software.
\textit{Mininet} uses \textit{process-based virtualization} and \textit{network namespaces} to run many%
\footnote{the developers successfully booted up to 4096}
hosts and switches on a single OS kernel. The network connections are realized through \textit{virtual ethernet (veth)} pairs and \textit{pipes}. \textit{Mininet} out-of-the-box supports kernel and user-space \textit{OpenFlow} switches and also controllers to control the switches.
\textit{Mininet} is very light-weight compared to many other virtualisation workflows. As shown in \cite{a19-lantz},
it can yield an efficient use of time and resources, providing interactive prototyping, scalability and straightforward sharing and collaboration for close-to-real-world developing and testing. For controlling the emulation setup, topology and conducted actions \textit{Mininet} offers on one side an extensible \textit{Python} API through which predefined actions, like network creation can be done. On the other hand there is also a command-line interface that allows controlling the environment as well as every single device live during runtime.

Due to the fact every \textit{Mininet} host can be assumed as being a full-featured Linux-device with ordinary access, the actual development topic does not necessarily have to be integrated in the emulation environment but rather can be independently worked on and just be \textit{called} from within the emulated network, exactly how using a real network. Furthermore this applies also for concurrent access of several developers.

Since \textit{Mininet} runs on Linux, uses Linux utilities, acts like Linux and does not aim to mimic any physical effects, it does not span any unique calculation basis and thus doesn't create an own isolated virtual time-base ("in it's own isolated world") that is depending on the computational model or something like an internal Event-Handling. Rather it acts like a real network would, only the scalability of an emulated network depending on the computation power of the executing device. Concomitant is \textit{Mininet} capable of interaction with real world devices. Using \acp{nic} of the emulation executing device packets can be exchanged between virtual \textit{Mininet} devices and the real world.

Whereas \textit{Mininets} goal is to provide a network to work on, i.e. to emulate the behaviour of an interconnection and not to simulate the behaviour of a physical transmission, it does not work on any sophisticated channel or loss model. \textit{Mininet} does indeed supply some parameters for the transmission connection to deliminate them from a perfect channel. For example one can limit the bandwidth or specify a statistical packet loss rate ($x\%$ packets lost) but that is not near as realistic as the opportunities made possible by low level network simulations. This fact essentially limits the representativity and accuracy of timing measurements for transmissions, when intended to work on lower layer solutions.

From its mentioned intention, \textit{Mininet} is highly usable for functional development like application level communication or timing-uncritical \ac{sdn}-algorithms but cannot suit performance related research that approaches how Data actually is transferred.

\section{Solution Mapping}
\label{sec:sol_map}
In this section,
all solutions presented in section~\ref{sec:solution} are mapped onto the requirements as derived in section~\ref{sec:req},
with notes on simulation and emulation capabilites,
as well as the scope that can be met.
An additional requirement,
ease of use,
has been added,
since commonly,
usability is an important feature in industrial applications.
This mapping is shown in table~\ref{tab:req_mapping}.

Solution \textit{S1},
application specific solutions,
provide the highest possible degrees of freedom for timing precision,
real world interaction and adaptability.
It is,
however,
neither flexible,
as hardcoded solutions are usually hardly adjustable,
nor easy to use,
respectively implement.
As it strongly depends on the singular case,
it can be both simulation as well as emulation,
and it can be placed on any layer of automation.

Solution \textit{S2}, which is \textit{ns3}, supports the use of a real-time scheduler which leads to a high achievable timing precision. This ability in conjunction with many models which are realistic enough to be used together with real-world network devices enables ns3 to be used as a real-time network emulator. Due to its open-source nature and the availability of many adaptable modules, the flexibility and adaptability of ns3 is high, although working in the ns3 core modules is effortful. The flexibility of ns3 arises also from its abstraction layer (e.g. node, net device, application, etc.) which allows the adaptation of the simulation on different layers. The physical interaction, in particular when it comes to wireless channels, is simulated using different models. The suitability and accuracy for different environments and use-cases follows from the accuracy of the underlying ns3 model, which is sufficient for a large number but not all of the imaginable situations. 

Solution \textit{S3},
\textit{OMNeT++},
allows for good timing precision,
is flexible and easy to adjust,
as well as easy to use.
The physical interaction on the other hand depends on the simulation model,
with all the drawbacks of artifacts and inconsistencies.
Furthermore,
it is only easily adaptable to scenarios and models already implemented in libraries,
novel methods take more implementation effort.
It is a simulation environment,
however,
it could be used as an emulation tool.
Physical sensor simulation is not possible,
as well as \ac{erp} and \ac{mes} are out of scope.

Solution \textit{S4}, \textit{Mininet}, 
does not possess an own time-base and its latency is limited by the resources provided by the host computer; deterministic timing is therefore not possible. It behaves like a Linux environment and its flexibility and adaptability is thus limited to the Linux world. Being an emulator, the interaction with real-world network devices is possible and commonplace. The consideration of physical aspects is limited to a few very generic possibilities to model the influence of a channel.

Additionally,
it should be remarked that cost and licensing has not been considered explicitly,
as it is not a functional requirement.
Emulation in general is more costly than simulation,
due to the hardware needed.
Furthermore,
while \textit{ns-3} and \textit{Mininet} are open source,
\textit{OMNeT++} demands for a license in case of industrial application.
On the other hand,
open source software usually contains so-called ``copyleft'',
forcing users to make tools using the software available.
\definecolor{geilesrot}{rgb}{ .753,  0,  0}
\definecolor{angenehmesorange}{rgb}{ 1,  .753,  0}
\definecolor{superduftesgruen}{rgb}{ 0,  .69,  .314}
\definecolor{woDieSonneNieHinscheintSchwarz}{rgb}{ 0,  0,  0}
	\newcommand{\legcrossstraight}{\textcolor{geilesrot}{$\times$}}
\newcommand{\legcross}{\textbf{\huge\legcrossstraight}}
	\newcommand{\legbulletstraight}{\textcolor{angenehmesorange}{$\bullet$}}
\newcommand{\legbullet}{\textbf{\huge\legbulletstraight}}
	\newcommand{\legcheckstraight}{\textcolor{superduftesgruen}{$\checkmark$}}
\newcommand{\legcheck}{\textbf{\huge\legcheckstraight}}

\begin{table}[!t]
\small
  \begin{center}
  \caption{Requirements Mapped to Solutions}
  \label{tab:req_mapping}
    \begin{tabular}{|l|c|c|c|c|}
		\hline
        \rowcolor{Gray}
		\begin{tabular}[t]{@{}lr@{}}
           & \textbf{Solution:} \\
    \textbf{Requirement:} &
  \end{tabular} & \textit{S1} & \textit{S2} & \textit{S3} & \textit{S4} \\
	\hline
	Timing Precision & \legcheck & \legcheck & \legcheck  & \legbullet\\
	\hline
	Flexibility & \legcross & \legcheck & \legcheck & \legbullet\\
	\hline
	Real World Interaction & \legcheck & \legcheck & \legcheck & \legcheck\\
	\hline
	Trusty Physical Interaction  & \legcheck & \legbullet & \legbullet & \legcross\\
	\hline
	Adaptability & \legcheck & \legcheck & \legbullet & \legbullet\\
	\hline
    Ease of Use & \legcross & \legbullet & \legcheck & \legbullet\\
	\hline
    Simulation/Emulation & Both & Both & Sim & Emu\\
	\hline
    Levels of Automation & 1-5 & 1-5 & 2-3 & 3-5\\
	\hline
		\end{tabular}%
		\end{center}
	\par
\makeatletter
    \footnotesize%
\makeatother
\end{table}


\section{Conclusion}
\label{sec:conclusion}
In this work,
four different kinds of scenarios in industrial applications are described.
The requirements posed on each of those scenarios are derived and summarised.
After that, 
commonly used solutions for network simulation are presented.
The features they provide are mapped against the requirements,
showing that most requirements can be met by one or another solution approach,
even though there is no general solution meeting all requirements.
The employment of a specific solution approach strongly depends on the use case and the kind of data needed.
In addition to that,
wireless networks require different simulation environments than wired networks,
and locally small networks have a different scope than widespread networks with heterogeneous entities.
While real world hard- and software provide the most realistic results, 
they are most difficult to set up,
often negating the possible advantage with the effort needed.

\section*{Acknowledgment}
This work has been supported by the Federal Ministry of Education and Research (BMBF) of the Federal Republic of Germany
within the projects IUNO (KIS4ITS0001) \& FIND (16KIS0571).
The authors alone are responsible for the content of the paper.

\bibliographystyle{unsrt}
\bibliography{literature}

\begin{thebibliography}{10}

\bibitem{karrenbauer2018}
Michael Karrenbauer, Amina Fellan, Hans~D Schotten, Henning Buhr, Savita
  Seetaraman, Norbert Niebert, Stephan Ludwig, Anne Bernardy, Vasco Seelmann,
  Volker Stich, et~al.
\newblock Towards a flexible architecture for industrial networking.
\newblock {\em 23th VDE/ITG Conference on Mobile Communication (23. VDE/ITG
  Fachtagung Mobilkommunikation), Osnabrück}, 2018.

\bibitem{Terzi.2004}
Segio Terzi and Sergio Cavalieri.
\newblock Simulation in the supply chain context: a survey.
\newblock {\em Computers in Industry}, 53(1):3 -- 16, 2004.

\bibitem{Weingartner.2009}
Elias Weingartner, Hendrick vom Lehn, and Klaus Wehrle.
\newblock A\ performance comparison of recent network simulators.
\newblock In {\em 2009 IEEE International Conference on Communications}, pages
  1--5, June 2009.

\bibitem{Flores_lucio.2003}
Gilberto Flores~Lucio, Marcos Paredes-Farrera, Emmanuel Jammeh, Martin Fleury,
  and Martin~J. Reed.
\newblock Opnet modeler and ns-2: Comparing the accuracy of network simulators
  for packet-level analysis using a network testbed.
\newblock {\em WSEAS Transactions on Computers}, 3(2):700--707, July 2003.

\bibitem{Pan.2008}
Jianli Pan.
\newblock A\ survey of network simulation tools: Current status and future
  developments, 2008.

\bibitem{Siraj.2012}
Saba Siraj, Ajay~Kumar Gupta, and Rinku-Badgujar.
\newblock Network simulation tools survey.
\newblock {\em International Journal of Advanced Research in Computer and
  Communication Engineering}, 1(4):201--210, June 2012.

\bibitem{Gokturk.2007}
Erek G\"{o}kt\"{u}rk.
\newblock A\ stance on emulation and testbeds, and a survey of network
  emulators and testbeds.
\newblock In {\em Proceedings 21st European Conference on Modelling and
  Simulation}, 2007.

\bibitem{Koksal.2008}
Murat~Miran K\"{o}sal.
\newblock A\ survey of network simulators supporting wireless networks, October
  2008.

\bibitem{Naicken.2006}
Stephen Naicken, Anirban Basu, Barnaby Livingston, and Sethalat Rodhetbhai.
\newblock A\ survey of peer-to-peer network simulators.
\newblock {\em PG NET}, 2006.

\bibitem{Martinez.2011}
Francisco~J. Martinez, Chai~Keong Toh, Juan-Carlos Cano, Carlos~T. Calafate,
  and Pietro Manzoni.
\newblock A\ survey and comparative study of simulators for vehicular ad hoc
  networks (vanets).
\newblock {\em Special Issue: Emerging Techniques for Wireless Vehicular
  Communications}, 11(7):813--828, July 2011.

\bibitem{Sarkar.2011}
Nurul~I. Sarkar and Syafnidar~A. Halim.
\newblock A\ review of simulation of telecommunication networks: Simulators,
  classification, comparison, methodologies and recommendations.
\newblock {\em Cyber Journals: Multidisciplinary Journals in Science and
  Technology, Journal of Selected Areas in Telecommunications (JSAT)}, pages
  10--17, March 2011.

\bibitem{Sundani.2011}
Haoyue Sundani, Vijay~K. Li, Mansoor Devabhaktuni, and Prabir Bhattacharya.
\newblock A\ review of simulation of telecommunication networks: Simulators,
  classification, comparison, methodologies and recommendations.
\newblock {\em International Journal Of Computer Networks (IJCN)},
  2(5):249--265, Janurary 2011.

\bibitem{Imran.2010}
Muhammad Imran, Abas~Md Said, and Halabi Hasbullah.
\newblock A\ survey of simulators, emulators and testbeds for wireless sensor
  networks.
\newblock In {\em 2010 International Symposium on Information Technology},
  volume~2, pages 897--902, 2010.

\bibitem{Korkalainen.2009}
Marko Korkalainen, Mikko Sallinen, Niilo Kärkkäinen, and Pirkka Tukeva.
\newblock Survey of wireless sensor networks simulation tools for demanding
  applications.
\newblock In {\em 2009 Fifth International Conference on Networking and
  Services}, pages 102--106, April 2009.

\bibitem{Christin.2010}
Delphine Christin, Parag~S. Mogre, and Matthias Hollick.
\newblock Survey on wireless sensor network technologies for industrial
  automation: The security and quality of service perspectives.
\newblock {\em future internet}, 2(2):96--125, April 2010.

\bibitem{Yu.2011}
Fei Yu.
\newblock A\ survey of wireless sensor network simulation tools, 2011.

\bibitem{3gpp2017}
Study on communication for automation in vertical domains, 2017.
\newblock 3GPP TR 22.804, V1.0.0.

\bibitem{DenKrChFiAR2018}
Santiago Soler~Perez Olaya, Martin Wollschlaeger, Dennis Krummacker, Christoph
  Fischer, Hans~D. Schotten, René Guillaume, Joachim~W. Walewski, and Norman
  Franchi.
\newblock Communication abstraction supports network resource virtualisation in
  automation.
\newblock {\em IEEE International Symposium on Industrial Electronics (ISIE)},
  27, 2018.

\bibitem{STEF2013451}
Ioan~Dorian Stef, George Draghici, and Anca Draghici.
\newblock Product design process model in the digital factory context.
\newblock {\em Procedia Technology}, 9:451 -- 462, 2013.

\bibitem{Fraunholz.2018b}
Daniel Fraunholz and Hans~Dieter Schotten.
\newblock Defending web servers with feints, distraction and obfuscation.
\newblock {\em International Conference on Computing, Networking and
  Communications}, 2018.

\bibitem{Fraunholz.2018}
Daniel Fraunholz, Daniel Krohmer, Frederic Pohl, and Hans~Dieter Schotten.
\newblock On the detection and handling of security incidents and perimeter
  breaches - a modular and flexible honeytoken based framework.
\newblock {\em Proceedings of the 9th IFIP International Conference on New
  Technologies, Mobility {\&} Security}, 9, 2018.

\bibitem{Fraunholz.2017e}
Daniel Fraunholz, Daniel Krohmer, Simon {Duque Anton}, and Hans~Dieter
  Schotten.
\newblock Investigation of cyber crime conducted by abusing weak or default
  passwords with a medium interaction honeypot.
\newblock {\em International Conference On Cyber Security And Protection Of
  Digital Services}, 2017.

\bibitem{Fraunholz.2017d}
Daniel Fraunholz, Simon {Duque Anton}, and Hans~Dieter Schotten.
\newblock Introducing gamfis: A generic attacker model for information
  security.
\newblock {\em International Conference on Software, Telecommunications and
  Computer Networks}, 25, 2017.

\bibitem{Fraunholz.2017f}
Daniel Fraunholz, Daniel Krohmer, Simon {Duque Anton}, and Hans~Dieter
  Schotten.
\newblock Yaas - on the attribution of honeypot data.
\newblock {\em International Journal on Cyber Situational Awareness},
  2(1):31--48, 2017.

\bibitem{Rist.2015}
Lukas Rist, Johnny Vestergaard, Daniel Haslinger, Andrea Pasquale, and John
  Smith.
\newblock Conpot ics/scada honeypot, 2015.

\bibitem{Provos.2004}
N.~Provos.
\newblock A\ virtual honeypot framework.
\newblock {\em USENIX Security Symposium}, 2004.

\bibitem{CarnivoreProject.2009}
{Carnivore Project}.
\newblock Dionaea: A malware capturing honeypot, 2009.

\bibitem{Oosterhof.2014}
Michel Oosterhof.
\newblock Cowrie, 2014.

\bibitem{Kuman.2017}
Stipe Kuman, Stjepan Gros, and Miljenko Mikuc.
\newblock An experiment in using imunes and conpot to emulate honeypot control
  networks.
\newblock In {\em 2017 40th International Convention on Information and
  Communication Technology, Electronics and Microelectronics (MIPRO)}, pages
  1262--1268. IEEE, 2017.

\bibitem{Litchfield.2017}
Samuel Litchfield.
\newblock {\em HoneyPhy: A physics-aware CPS honeypot framework}.
\newblock PhD thesis, {Georgia Tech}, 2017.

\bibitem{GridPot:SymbolicCyberPhysicalHoneynetFramework.2015}
{GridPot: Symbolic Cyber-Physical Honeynet Framework}.
\newblock sk4ld, 2015.

\bibitem{Redwood.2015}
Owen Redwood, Joshua Lawrence, and Mike Burmester.
\newblock A symbolic honeynet framework for scada system threat intelligence.
\newblock In Mason Rice and Sujeet Shenoi, editors, {\em Critical
  Infrastructure Protection IX}, volume 466 of {\em IFIP Advances in
  Information and Communication Technology}, pages 103--118. {Springer
  International Publishing}, Cham, 2015.

\bibitem{Duque_Anton.2017a}
Simon Duque~Anton, Daniel Fraunholz, Christoph Lipps, Frederic Pohl, Marc
  Zimmermann, and Hans~Dieter Schotten.
\newblock Two decades of scada exploitation: A brief history.
\newblock In {\em IEEE Conference on Applications, Information and Network
  Security (AINS). IEEE Conference on Applications, Information and Network
  Security (AINS-2017), November 13-14, Miri, Sarawak, Malaysia}, 2017.

\bibitem{Duque_Anton.2017c}
Simon Duque~Anton, Daniel Fraunholz, Stephan Teuber, and Hans~Dieter Schotten.
\newblock A question of context: Enhancing intrusion detection by providing
  context information.
\newblock In {\em 13th Conference of Telecommunication, Media and Internet
  Techno-Economics (CTTE-17)}, 2017.

\bibitem{Duque_Anton.2017b}
Simon Duque~Anton, Daniel Fraunholz, Janis Zemitis, Frederic Pohl, and
  Hans~Dieter Schotten.
\newblock Highly scalable and flexible model for effective aggregation of
  context-based data in generic iiot scenarios.
\newblock In {\em 9th Central European Workshop on Services and their
  Composition. Central European Workshop on Services and their Composition
  (ZEUS-2017), February 13-14, Lugano, Switzerland}, pages 51--58, 4 2017.

\bibitem{Duque_Anton.2018}
Simon Duque~Anton, Suneetha Kanoor, Daniel Fraunholz, and Hans~Dieter Schotten.
\newblock Evaluation of machine learning-based anomaly detection algorithms on
  an industrial modbus/tcp data set.
\newblock In {\em Proceedings of the 13th International Conference on
  Availability, Reliability and Security}, 2018.

\bibitem{Garcia-Teodoro.2009}
Vinay~M. Igure, Sean~A. Laughter, and Ronald~D. Williams.
\newblock Anomaly-based network intrusion detection: Techniques, systems and
  challenges.
\newblock {\em Computers {\&} Security}, (28):18--28, February 2009.

\bibitem{Bhuyan.2014}
Monowar~H. Bhuyan, D.~K. Bhattacharyya, and J.~K. Kalita.
\newblock Network anomaly detection: Methods, systems and tools.
\newblock {\em IEEE Communications Surveys {\&} Tutorials}, 16(1):303--336,
  2014.

\bibitem{Yang.2006}
Dayu Yang, Alexander Usynin, and J.~Wesley Hines.
\newblock Anomaly-based intrusion detection for scada systems.
\newblock In {\em 5. International Topical Meeting on Nuclear Plant
  Instrumentation Controls, and Human Machine Interface Technology}, pages
  797--803, November 2006.

\bibitem{Meshram.2016}
Ankush Meshram and Christian Haas.
\newblock Anomaly detection in industrial networks using machine learning: A
  roadmap.
\newblock In {\em Machine Learning for Cyber Physical Systems}, pages 65--72,
  2016.

\bibitem{Kleinmann.2014}
Amit Kleinmann and Avishai Wool.
\newblock Accurate modeling of the siemens s7 scada protocol for intrusion
  detection and digital forensics.
\newblock In {\em Journal of Digital Forensics, Security and Law}, volume~9,
  2014.

\bibitem{Hadziosmanovic.2011}
Dina Hadziosmanovic, Damiano Bolzoni, Pieter Hartel, and Sandro Etalle.
\newblock {\em MELISSA: Towards Automated Detection of Undesirable User Actions
  in Critical Infrastructures}, pages 41--48.
\newblock 2011.

\bibitem{Mantere.2013}
Matti Mantere, Mirko Sailio, and Sami Noponen.
\newblock Network traffic features for anomaly detection in specific industrial
  control system network.
\newblock {\em Future Internet}, 4(5):460--473, 2013.

\bibitem{Hadeli.2009}
Hadeli Hadeli, Ragnar Schierholz, Markus Braendle, and Christian Tuduce.
\newblock Leveraging determinism in industrial control systems for advanced
  anomaly detection and reliable security configuration.
\newblock In {\em 2009 IEEE Conference on Emerging Technologies Factory
  Automation}, pages 1--8, 2009.

\bibitem{Morris.2014}
Thomas Morris and Wei Gao.
\newblock {\em Industrial Control System Traffic Data Sets for Intrusion
  Detection Research}, pages 65--78.
\newblock 2014.

\bibitem{Cellier.1991}
Francois~E. Cellier and Jurgen Greifeneder.
\newblock {\em Continuous System Modeling}.
\newblock 1st edition.

\bibitem{KB.}
Koninklijke Bibliotheek National~Library of~the Netherlands.
\newblock Emulation.

\bibitem{Fraunholz.2017g}
Daniel Fraunholz, Frederic Pohl, and Hans~Dieter Schotten.
\newblock Towards basic design principles for high- and medium-interaction
  honeypots.
\newblock {\em European Conference on Cyber Warfare and Security}, 16, 2017.

\bibitem{Fu.2006}
X.~Fu, W.~Yu, D.~Cheng, X.~Tan, and S.~Graham.
\newblock On recognizing virtual honeypots and countermeasures.
\newblock {\em International Symposium on Dependable, Autonomic and Secure
  Computing}, 2, 2006.

\bibitem{Chen.2008}
Xu~Chen, Jon Anderson, Morley Mao, Michael Bailey, and Jose Nazario.
\newblock Towards an understanding of anti-virtualization and anti-debugging
  behavior in modern malware.
\newblock {\em International Conference on Dependable Systems {\&}Networks},
  pages 177--186, 2008.

\bibitem{Mukkamala.2007}
S.~Mukkamala, K.~Yendrapalli, R.~Basnet, M.~Shankarapani, and H.~Sung.
\newblock Detection of virtual environments and low interaction honeypots.
\newblock {\em Information Assurance and Security Workshop}, 2007.

\bibitem{Dornseif.2004}
M.~Dornseif, T.~Holz, and Christian Klein.
\newblock Nosebreak - attacking honeynets.
\newblock {\em Workshop on Information Assurance and Security}, pages 123--129,
  2004.

\bibitem{Holz.2005}
T.~Holz and F.~Raynal.
\newblock Detecting honeypots and other suspicious environments.
\newblock {\em Workshop on Information Assurance and Security}, pages 1--8,
  2005.

\bibitem{Fraunholz.2017b}
Daniel Fraunholz, Marc Zimmermann, and Hans~Dieter Schotten.
\newblock An adaptive honeypot configuration, deployment and maintenance
  strategy.
\newblock {\em International Conference on Advanced Communications Technology},
  19, 2017.

\bibitem{Fraunholz.2017}
Daniel Fraunholz, Marc Zimmermann, and Hans~Dieter Schotten.
\newblock Towards deployment strategies for deception systems.
\newblock {\em Advances in Science, Technology and Engineering Systems
  Journal}, Special Issue on Recent Advances in Engineering Systems, 2017.

\bibitem{Fraunholz.2017i}
Daniel Fraunholz, Christoph Lipps, Marc Zimmermann, Simon {Duque Anton}, and
  Hans~Dieter Schotten.
\newblock Deception in information security: Legal considerations in the
  context of german and european law.
\newblock {\em International Symposium on Foundations {\&} Practice of
  Security}, 10, 2017.

\bibitem{Lemay.2016}
Antoine Lemay and Jose~M. Fernandez.
\newblock Providing scada network data sets for intrusion detection research.
\newblock In {\em 9th Workshop on Cyber Security Experimentation and Test (CSET
  16)}, Austin, TX, 2016.

\bibitem{Wang.2010}
Chunlei Wang, Lan Fang, and Yiqi Dai.
\newblock A simulation environment for scada security analysis and assessment.
\newblock In {\em 2010 International Conference on Measuring Technology and
  Mechatronics Automation}, volume~1, pages 342--347, March 2010.

\bibitem{Seidl.2015}
Jan Seidl.
\newblock Virtuaplant.

\bibitem{Siaterlis.2013}
Christos Siaterlis, Bela Genge, and Marc Hohenadel.
\newblock Epic: A testbed for scientifically rigorous cyber-physical security
  experimentation.
\newblock {\em IEEE Transactions on Emerging Topics in Computing},
  1(2):319--330, 2013.

\bibitem{Genge.2012}
Bela Genge, Christos Siaterlis, Igor~Nai Fovino, and Marcelo Masera.
\newblock A cyber-physical experimentation environment for the security
  analysis of networked industrial control systems.
\newblock {\em Computers {\&} Electrical Engineering}, 38(5):1146--1161, 2012.

\bibitem{boeingNS3ChanEval}
Guangyu Pei and Thomas~R. Henderson.
\newblock {\em Validation of OFDM error rate model in ns-3}.
\newblock Boeing Research \& Technology, 2010.

\bibitem{OMNeT++.2015}
OMNeT++.
\newblock Omnet++ discrete event simulator, 2015.

\bibitem{mininetWeb}
Mininet.

\bibitem{a19-lantz}
Bob Lantz, Brandon Heller, and Nick McKeown.
\newblock A\ network in a laptop: Rapid prototyping for software-defined
  networks.
\newblock Paper, Stanford University and DOCOMO USA Labs, 2000.

\end{thebibliography}

\end{document}